\documentclass[a4paper]{jpconf}
\usepackage{graphicx}
\usepackage{amssymb}
\usepackage{amsmath}
\usepackage{amsfonts}
\usepackage{mathtools}

\usepackage{cite}
\usepackage{ulem}
\usepackage[dvipsnames,svgnames]{xcolor}

\begin{document}
\title{Photon-Photon Scattering at the High-Intensity Frontier: Paraxial Beams}

\author{A. Blinne$^{1}$, H. Gies$^{1,2,3}$, F. Karbstein$^{1,2}$, C. Kohlf\"urst$^{1,2}$ and M. Zepf$^{1,4}$}

\address{$^1$Helmholtz Institut Jena, Jena, Germany}
\address{$^2$Theoretisch-Physikalisches Institut, Friedrich-Schiller-Universit\"at Jena, Jena, Germany}
\address{$^3$Abbe Center of Photonics, Jena, Germany}
\address{$^4$Institute of Optics and Quantum Electronics, Friedrich-Schiller-Universit\"at Jena, Jena, Germany}

\ead{c.kohlfuerst@gsi.de}

\begin{abstract}
  Our goal is to study optical signatures of quantum vacuum nonlinearities in strong  macroscopic electromagnetic fields provided by high-intensity laser beams.  
  The vacuum emission scheme is perfectly suited for this task as it naturally distinguishes between incident laser beams, described as classical electromagnetic fields driving the effect,
  and emitted signal photons encoding the signature of quantum vacuum nonlinearity.
  Using the Heisenberg-Euler effective action, our approach allows for a reliable study of photonic signatures of QED vacuum nonlinearity in the parameter regimes accessible by all-optical high-intensity laser experiments.
  To this end, we employ an efficient, flexible numerical algorithm, which allows for a detailed study of the signal photons
  emerging in the collision of focused paraxial high-intensity laser pulses. Due to the high accuracy of our numerical solutions
  we predict the total number of signal photons, but also have full access to the signal photons' characteristics,
  including their spectrum, propagation directions and polarizations. We discuss setups
  offering an excellent background-to-noise ratio, thus providing an important step towards the experimental verification of quantum vacuum nonlinearities.
\end{abstract}

\section{Introduction}

The effective interactions of electromagnetic fields in the vacuum are one of the most surprising predictions of strong-field 
quantum electrodynamics (QED) \cite{Euler:1935zz,Heisenberg:1935qt,Weisskopf,Schwinger:1951nm}.
They are induced by vacuum fluctuations and supplement Maxwell's linear theory of electrodynamics with nonlinear corrections.
Even though QED vacuum nonlinearities have already been predicted theoretically more than 80 years ago, so far they have never been verified in macroscopic electromagnetic fields.
The reason is that they are parametrically suppressed with
the electron mass $m_e \approx 511$keV for electromagnetic fields of optical up to X-ray frequencies $\omega\ll m_ec^2/\hbar$, translating into a so-called critical electric $E_{\rm cr}=m_e^2c^3/(e\hbar) \approx 1.3 \times 10^{16}$V/cm 
and critical magnetic $B_{\rm cr}=E_{\rm cr}/c \approx 4 \times 10^{9}$T reference field strength. 
These critical fields are much larger than the strengths of all macroscopic electric $E$ and magnetic $B$ fields presently available in the laboratory.
In comparison to classical Maxwell theory, the leading QED vacuum nonlinearity is suppressed by factors of $(E/E_{\rm cr})^{2-n}(B/B_{\rm cr})^n$, with $n\in\{0,1,2\}$.

Only due to the recent progress in laser technology, which enabled high-intensity laser systems to reach peak electric and magnetic fields of the order of $E={\cal O}(10^{12}){\rm V}/{\rm cm}$ and $B={\cal O}(10^6){\rm T}$, respectively, the experimental study of QED vacuum nonlinearities has come within reach. 
Prominent examples of laser facilities at the high-intensity frontier are CILEX \cite{CILEX}, CoReLS \cite{CoReLS}, ELI \cite{ELI} and
SG-II \cite{SG-II}. At the same time, the accurate theoretical description
of photonic signatures of quantum vacuum nonlinearity in focused high-intensity laser pulses has substantially improved \cite{Marklund:2008gj,Heinzl:2008an,DiPiazza:2011tq,King:2015tba,Karbstein:2016hlj}. 
To be more specific, the typical challenge amounts to isolating a few signal photons from the background of a huge number ${\cal O}(10^{20})$ of laser photons driving the effect. 
To this end, it is absolutely essential to identify scenarios giving rise to signal photons with distinct propagation and polarization properties, allowing for a clear signal-to-background separation. 
A favorable scenario is given by signal photons propagating outside the forward cones of the driving laser beams scattered into a perpendicularly polarized mode.  
All in all, combining high-intensity
lasers, which provide huge peak field strengths to drive the effect, with polarization and direction sensitive single-photon detectors seems to be a promising route for detecting QED vacuum non-linearities for the first time.

In this article, we study photonic signatures of QED vacuum nonlinearities in high-intensity laser pulse collisions based on the ``vacuum emission picture'' \cite{Karbstein:2014fva}. This theoretical description closely parallels the typical experimental scenario: The signal photons are induced in the tiny interaction region, where the focused high-intensity laser pulses collide, and are detected far outside of it. 
Describing the driving laser fields as pulsed paraxial beams, we can consider generic collision geometries of the driving laser pulses and
search the parameter space in order to identify a configuration, where the emitted photons differ significantly from the incident laser photons, e.g. by their kinematics.
This allows us to theoretically isolate optimal scenarios where quantum vacuum nonlinearities can become detectable in experiment. 

This article is organized as follows: after briefly reviewing the formalism in Sec. \ref{sec:formalism} and the adopted beam model in Sec.~\ref{sec:beams}, we discuss some exemplary numerical results in Sec. \ref{sec:results}. 
Finally, we conclude with a summary and an outlook in Sec. \ref{sec:summary}.
Throughout the article we use the Heaviside-Lorentz system with $c=\hbar=1$ and the metric $g_{\mu \nu} = {\rm diag} (-1,+1,+1,+1)$.

\section{Formalism}
\label{sec:formalism}

A complete derivation of the vacuum emission formalism is given in Refs. \cite{Karbstein:2014fva,Gies:2017ezf,Gies:2017ygp}. 
In order to keep this article self-contained, here we briefly recall the key elements of the approach. For alternative methods in a similar photon-scattering context, see, e.g., Refs. \cite{Moulin:2002ya,Bernard:2010dx,Tommasini:2010fb,Paredes:2014oxa,King:2012aw,Lundstrom:2005za,Lundin:2006wu,King:2018wtn}.

At leading order in a loop expansion, the zero-to-one signal photon emission amplitude is given by
\begin{equation}
 {\cal S}_{(p)}(\vec{k})\equiv\big\langle\gamma_p(\vec{k})\big| \Gamma_\text{int}^{(1)}[A(x)] \big|0\big\rangle \, ,
\end{equation}
where $|\gamma_{p}(\vec{k})\rangle\equiv a^\dag_{\vec{k},p}|0\rangle$ denotes the single signal photon state with wave vector $\vec{k}$ and polarization $p$, 
$\Gamma_\text{int}^{(1)}[A(x)]$ is the one-loop Euler-Heisenberg effective action and
$A$ denotes the gauge potential of the macroscopic electromagnetic field driving the effect. 
Resorting to the local constant field approximation (LCFA) of the Heisenberg-Euler effective Lagrangian ${\cal L}_{\rm HE}(F)$ in constant electromagnetic fields $F$ \cite{Heisenberg:1935qt}, this expression can be recast as
\begin{equation}
{\cal S}_{(p)}(\vec{k})
 =\frac{\epsilon^{*\nu}_{(p)}(\vec{k})}{\sqrt{2 {\rm k}}} \, 2{\rm i} k^\mu\int{\rm d}^4 x\,{\rm e}^{{\rm i}kx}\,\frac{\partial{\cal L}_\text{HE}^{1\text{-loop}}(F)}{\partial F^{\mu\nu}}\bigg|_{F\to F(x)}\,, \label{eq:Sp}
\end{equation}
with $\epsilon^{*\nu}_{(p)}(\vec{k})$ representing the polarization vector of the signal photon and $k^0=|\vec{k}|={\rm k}=\sqrt{k_x^2 + k_y^2 + k_z^2}$.
Using spherical coordinates $\vec{k}={\rm k}\hat{\vec{e}}_k$ we can parametrize the signal photon's propagation direction as
\begin{equation}
 \hat{\vec{e}}_k=
\left(\begin{array}{c}
  \cos\varphi\sin\vartheta \\
  \sin\varphi\sin\vartheta \\
  \cos\vartheta
 \end{array}\right) ,
\end{equation}
and the unit vectors perpendicular to it as
\begin{equation}
\hat{\vec{e}}_{\beta}=
\left(\begin{array}{c}
  \cos\varphi\cos\vartheta\cos\beta-\sin\varphi\sin\beta \\
  \sin\varphi\cos\vartheta\cos\beta+\cos\varphi\sin\beta \\
  -\sin\vartheta\cos\beta
 \end{array}\right) . 
\end{equation}
The latter can be employed to span
the polarization basis of signal photons of wave vector $\vec{k}$, defining $\epsilon^\mu_{(p)}(\vec{k}):=(0,\hat{\vec{e}}_{\beta=\beta_0+\frac{\pi}{2}(p-1)})$ with $p\in\{1,2\}$ and
$\beta_0 \in \mathbb{R}$.

The leading contribution in a perturbative weak-field expansion $e F^{\mu \nu} \ll m_e^2$ of eq. \eqref{eq:Sp}, is given by
\begin{align}
&{\cal S}_{(p)}(\vec{k})
 = \frac{1}{\rm i}\frac{e}{4\pi^2}\frac{m_e^2}{45}\sqrt{\frac{\rm k}{2}}\Bigl(\frac{e}{m_e^2}\Bigr)^3 \int{\rm d}^4 x\, {\rm e}^{{\rm i}{\rm k}(\hat{\vec{e}}_k\cdot\vec{x}-t)} \nonumber\\
&\,\quad\times\Bigl\{4\bigl[\hat{\vec{e}}_{\beta_0+\frac{\pi}{2}(p-1)}\cdot\vec{E}(x)-\hat{\vec{e}}_{\beta_0+\frac{\pi}{2}p}\cdot\vec{B}(x)\bigr] {\cal F}(x) \nonumber
+7\bigl[\hat{\vec{e}}_{\beta_0+\frac{\pi}{2}(p-1)}\cdot\vec{B}(x)+\hat{\vec{e}}_{\beta_0+\frac{\pi}{2}p}\cdot\vec{E}(x)\bigr] {\cal G}(x)\Bigr\},
\end{align}
with ${\cal F}(x)=\frac{1}{2}[\vec{B}^2(x)-\vec{E}^2(x)]$ and ${\cal G}(x)=-\vec{B}(x)\cdot\vec{E}(x)$.
Eventually, the differential number of signal photons in spherical coordinates takes the form
\begin{equation}
{\rm d}^3N_{(p)}(\vec{k})={\rm dk}\,{\rm d}\varphi\,{\rm d}\!\cos\vartheta\,\frac{1}{(2\pi)^3}\bigl|{\rm k}{\cal 
S}_{(p)}(\vec{k})\bigr|^2 \,. 
\label{eq:dN}
\end{equation}

\section{Model for the beams}
\label{sec:beams}

Throughout this article we rely on an analytical model of the driving high-intensity laser beams. For definiteness, we 
describe them as pulsed, linearly polarized, paraxial Gaussian beams. Without loss of generality, the field profile of a beam propagating along the positive z axis and focused at ${\rm z}={\rm t}={\rm r}=0$ is given by
\begin{equation}
 {\cal E}(x) = {\cal E}_{0}\,{\rm e}^{-\frac{({\rm z}-{\rm t})^2}{(\tau/2)^2}}\, \frac{w_{0}}{w({\rm z})}\, {\rm e}^{-\frac{r^2}{w^2({\rm z})}}\, \cos\bigl(\Phi(x)\bigr) \,,
 \label{equ:E}
\end{equation}
where 
\begin{equation}
 \Phi(x) = \omega({\rm z}-{\rm t})+\tfrac{{\rm z}}{{\rm z}_{R}}\tfrac{r^2}{w^2({\rm z})}-\arctan\bigl(\tfrac{{\rm z}}{{\rm z}_{R}}\bigr) \,.
  \label{equ:P}
\end{equation}
Here, $\omega=2\pi/\lambda$ is the laser frequency, $\tau$ the pulse duration and $w_0=\rho \lambda$ with $\rho > 0$ the beam waist. 
The peak field strength ${\cal E}_{0}$ can be related to the laser pulse energy $W$ as ${\cal E}_{0}^2\approx8\sqrt{\frac{2}{\pi}}\frac{W}{\pi w_{0}^2\tau}$ \cite{Karbstein:2017jgh}.
Furthermore, the Rayleigh range ${\rm z}_{{\rm R}}=\pi w_{0}^2/\lambda$ is the distance from the focus where the transverse section of the beam is increased by a factor of two, and 
$w({\rm z}) = w_{0} \sqrt{1+({\rm z}/{\rm z}_{{\rm R}})^2}$. The curvature of the wavefronts is described by the term $\tfrac{{\rm z}}{{\rm z}_{R}}\tfrac{r^2}{w^2({\rm z})}$,
while $\arctan({\rm z}/{\rm z}_{R})$ is the Gouy phase shift. As a function of the longitudinal coordinate z from the beam focus, the transverse section of a
Gaussian beam increases, its radial divergence in the far field is given by $\theta\simeq w_{0}/{\rm z}_{{\rm R}}$.

At leading order in the paraxial approximation, the electric and magnetic field vectors characterizing a given laser pulse are given by
\begin{equation}
 \vec{E}(x)={\cal E}(x)\hat{\vec{e}}_{E}, \ 
 \vec{B}(x)={\cal E}(x)\hat{\vec{e}}_{B}. 
 \label{eq:vecs}
\end{equation}
With $\hat{\vec{e}}_{\kappa}$ denoting the beam's propagation direction and $\hat{\vec{e}}_{E}$ its polarization vector, we have $\hat{\vec{e}}_{\kappa}\cdot\hat{\vec{e}}_{E}=0$ and $\hat{\vec{e}}_{\kappa}\times\hat{\vec{e}}_{E}=\hat{\vec{e}}_{B}$. Note, that upon identification of 
${\rm t}:=t-t_{0}$, ${\rm z}:=\hat{\vec e}_{\kappa}\cdot\left(\vec{x}-\vec{x}_{0}\right)$ and $r:=\sqrt{(\vec{x}-\vec{x}_{0})^2-{\rm z}^2}$ in eqs.~\eqref{equ:E}-\eqref{equ:P}, the above field profile can be straightforwardly invoked to describe a laser pulse propagating in direction $\hat{\vec{e}}_{\kappa}$ focused at $\vec{x}=\vec{x}_{0}$ and $t=t_{0}$. 

All results discussed in this article are based upon a numerical evaluation of eq.~\eqref{eq:dN} with the driving electromagnetic fields given by a superposition of laser pulses described by eqs.~\eqref{equ:E}-\eqref{eq:vecs}.
To this end, we have developed a numerical algorithm capable of handling any given collision scenario of high-intensity laser pulses which can be described as pulsed
paraxial beams. The technical details of the computation as well as a thorough analysis of the computational advantages of our approach are given in Ref. \cite{Gies:2017ygp}. 

\section{Results}
\label{sec:results}

In the following, we discuss the signal photon emission characteristics for two different experimental scenarios, envisioning the collision of two and three high-intensity laser pulses, respectively.
More specifically, we review the main results of our recent articles focusing on the collision of two \cite{Gies:2017ygp} and three \cite{Gies:2017ezf} laser beams and put them into a different perspective here. 

In order to provide realistic predictions of the numbers of attainable signal photons in all-optical laser experiments, we assume to have two identical high-intensity lasers of the one-petawatt class at our disposal rendering such an experiment possible at ELI-NP \cite{ELI}. 
More specifically, we assume these lasers to deliver pulses of energy $W=25$J with pulse duration $\tau=25$fs at a wavelength of $\lambda=800$nm, corresponding to a laser photon frequency of $\omega\approx1.55$eV.
The signal photon number is very sensitive to the fields' peak intensities in the interaction volume. In order to maximize these, we envision all beams to be focused down to the diffraction limit.
Furthermore, we assume optimal collisions and thus neglect any form of displacement or misalignment, $t_0=0$ and $\vec x_0=0$, of the beam foci. 

\subsection{Two-beam scenario}
\label{sec:2beams}

\begin{figure}[t]
\center
\includegraphics[width=12.0cm,trim=0 350 550 50, clip]{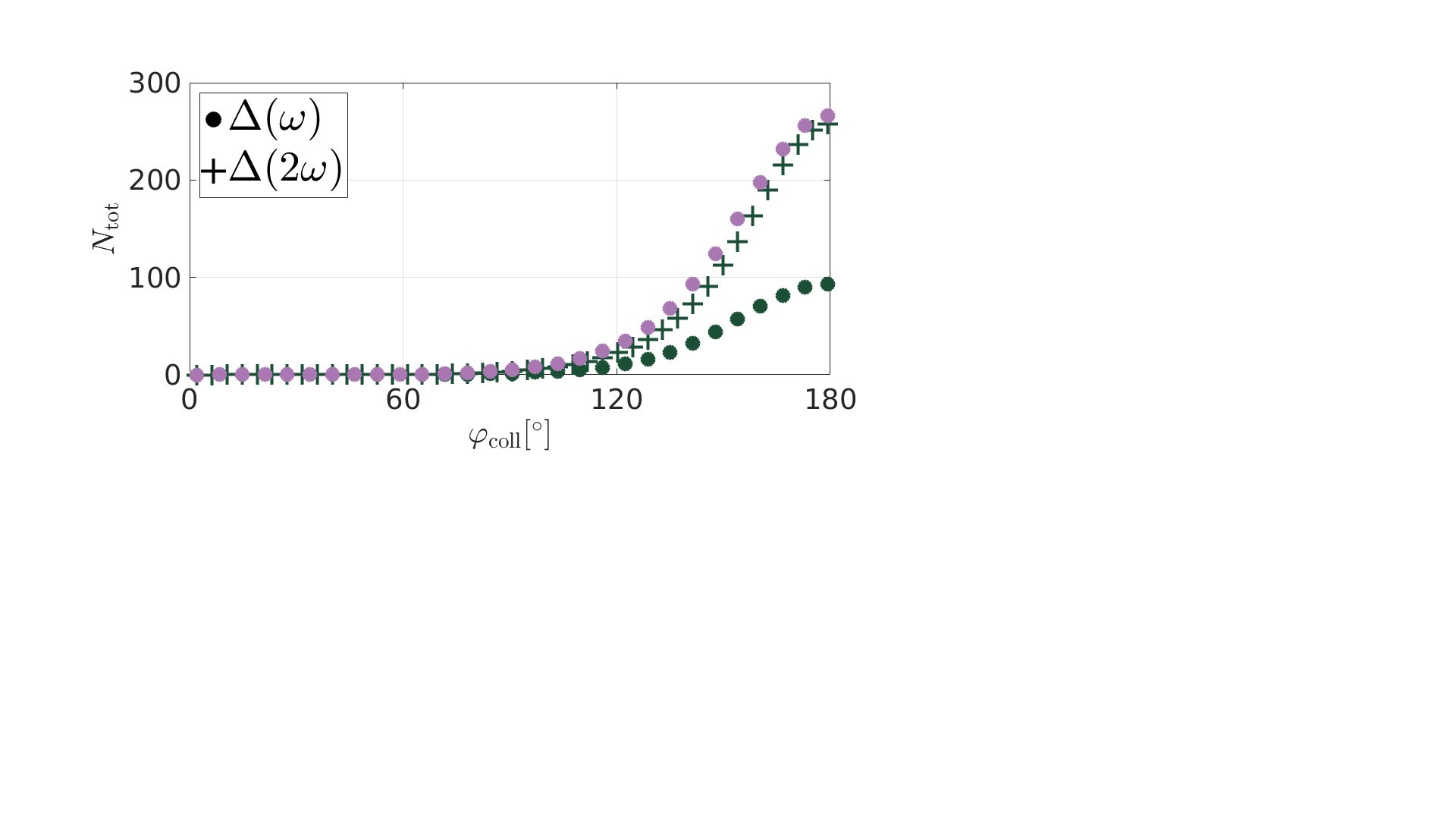} 
\caption{Integrated signal photon numbers partitioned into the energy regimes $\Delta(\omega)$ and $\Delta(2\omega)$ as a function of the collision angle $\varphi_{\rm coll}$ 
for the two-beam scenarios:
(i) $W_1=W_2$, $\omega_1=\omega_2$, $w_{0,1} = w_{0,2} = \lambda$ and (ii) $W_1=2W_2$, $\omega_2 = 2\omega_1$, $w_{0,1} = 2w_{0,2} = \lambda$. 
The magenta curve is for scenario (i) employing fundamental-frequency $\omega$ beams only, and the green curves are for scenario (ii) involving a frequency-doubled beam. 
In scenario (i), no
signal photons with energy $2 \omega$ are induced.}
\label{fig:Ntwo}
\end{figure}

Here, we focus on the behavior of the total number of emitted signal photons as a function of the collision angle $\varphi_{\rm coll}$. 
To this end, we consider two different scenarios. The most straightforward scenario assumes the collision of two identical laser pulses with pulse energy $W_1=W_2=W$, pulse duration $\tau_1=\tau_2=\tau$ and laser photon frequency $\omega_1=\omega_2=\omega$, focused to $w_{0,1}=w_{0,2}=\lambda$. 
In the alternative configuration, we assume the second laser pulse to be frequency doubled, such that $\omega_2 = 2\omega_1 = 2\omega$. Estimating an energy loss of $50\%$ for the frequency doubling process preserving the pulse duration, we have $W_1 = 2W_2 = W$ and
$\tau_1=\tau_2=\tau$. The beams are assumed to be maximally focused also in this case, i.e., $w_{0,1} = 2w_{0,2} = \lambda$. For definiteness, we assume the driving laser pulses 
to be polarized perpendicularly to the collision plane in both scenarios.

In Fig. \ref{fig:Ntwo} we illustrate the results for the number of signal photons for the two scenarios as a function of the collision angle $\varphi_{\rm coll}$.
Here, we distinguish between signal photons of energy $\omega$ and $2\omega$,
partitioned into the two different energy segments $\Delta (\omega)$ and $\Delta (2\omega)$, respectively.
The energies and propagation directions of the signal photons are determined by the kinematics of the effective interaction of three laser photons \cite{Gies:2017ygp}. 
For the scenario with $\omega_1=\omega_2=\omega$, this in particular constraints the possible signal photon energies as  
$\omega_{\ast} \approx \lvert \omega \pm \omega \pm \omega \rvert$.
Taking into account, that photon merging signals are strongly suppressed \cite{Yakovlev:1966,Karbstein:2014fva,Gies:2016czm}, we therefore obtain a signal photon frequency of $\omega_{\ast} \approx \omega$.
In the second scenario with $\omega_2=2\omega_1=2\omega$ we have two possibilities, namely either 
$\omega_{\ast} \approx \lvert \omega \pm 2\omega \pm 2\omega \rvert$ or $\omega_\ast\approx|2\omega\pm\omega\pm\omega|$. 
As photon merging processes are again highly suppressed we can expect to observe signal photon frequencies of $\omega_{\ast} = \omega$ and $\omega_{\ast} = 2\omega$ \cite{Gies:2017ygp}.

\subsection{Three-beam scenario}
\label{sec:3beams}

\begin{figure}[t]
\center
\includegraphics[width=7cm,trim=150 45 0 0, clip]{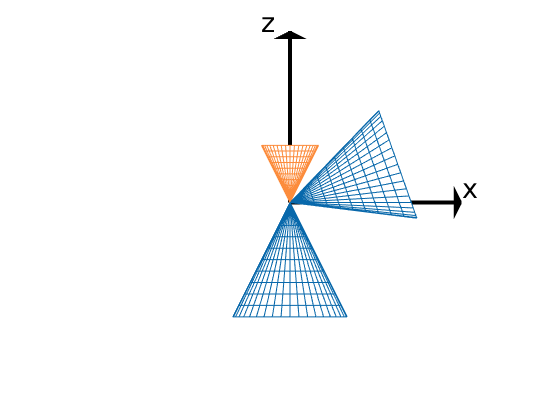}
\includegraphics[width=8cm,trim=265 0 275 40, clip]{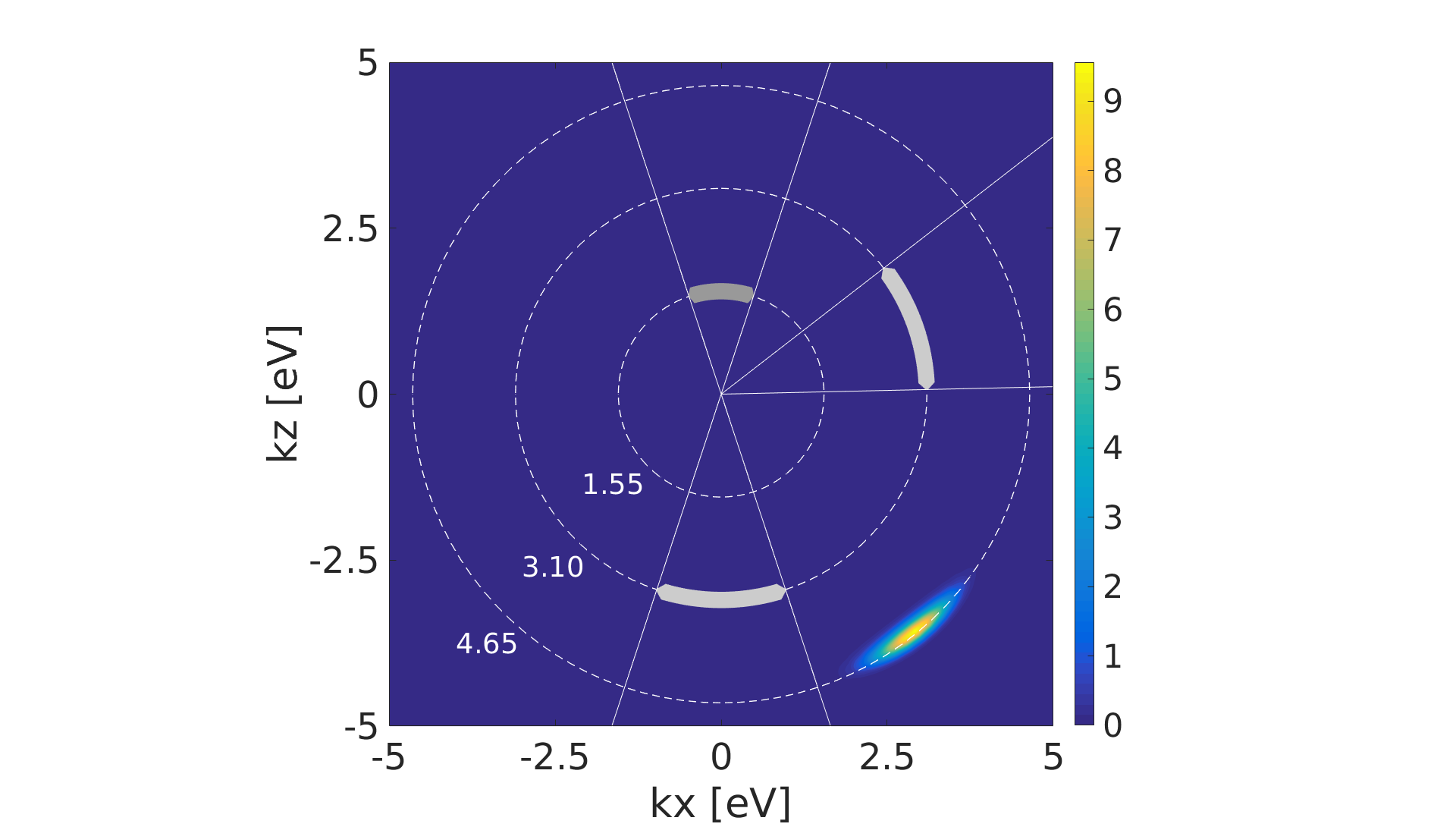}
\caption{Left: Sketch of a three-beam setup allowing for a clear signal-to-background separation. One fundamental-frequency $\omega_1 = \frac{2\pi}{\lambda} = 1.55$eV pulse (light orange) with energy $W_1=25\,{\rm J}$ is brought into collision with two frequency-doubled pulses (dark blue) with energy $W_2=W_3=6.25\,{\rm J}$ and frequency $\omega_2=\omega_3 = 3.10$eV.
The pulses have the same pulse duration $\tau_1=\tau_2=\tau_3=25$fs. All beams are focused to their diffraction limits, i.e., $w_{0,1}=2w_{0,2}=\lambda$. The cones represent the far-field divergences of the beams in forward direction. The beam axes of all beams are confined to the xz-plane
and exhibit collision angles of $\vartheta_2=180^\circ$ and $\vartheta_3=70.47^\circ$ with respect to the fundamental-frequency beam.

Right: Differential number $\frac{{\rm d}^2N_{\rm tot}}{{\rm d}k_x{\rm d}k_z}$ of signal photons attainable in the discussed three-beam setup.
  The signal photons have a distinct frequency of $\omega_{\ast} \approx 4.65$eV and are emitted in directions outside the forward cones of the high-intensity lasers, delimited by the white radial lines.
  The grey-colored areas mark the bulks of driving laser photons.}
\label{fig:sketch}
\end{figure}

The three-beam experiment we have in mind is the following: One of the lasers is kept as it is, and the second one is split equally into two parts, which are subsequently frequency-doubled. 
As above, we estimate the energy loss in the frequency doubling process as $50\%$. Therefore, we have $W_1 = 25$J and $W_2=W_3 = 6.25$J, $\tau_1 = \tau_2= \tau_3 = 25$fs as well as $\omega_1 =\omega= 1.55$eV and $\omega_2=\omega_3 = 2\omega= 3.1$eV.
The optimal collision geometry for the special case where the beam axes of all the driving laser pulses are confined to the xz-plane is depicted in Fig.~\ref{fig:sketch}; the second and third laser beams enter under
angles of $\vartheta_2=180^\circ$ and $\vartheta_3=70.47^\circ$ with respect to the first laser beam. For definiteness, we assume all beams to be polarized perpendicular to the propagation plane.
An in-depth analysis on the identification of ``optimal'' scenarios can be found in Ref. \cite{Seegert}. For effects of spatial and temporal displacements we refer to Ref.~\cite{Gies:2017ygp}.


For the specific collision scenario detailed here, we obtain $N_{\rm tot} \approx 3.03$ signal
photons per shot indicating that this setup could potentially even outperform the $3d$-scenario suggested in Ref. \cite{Lundstrom:2005za}, which yielded $N_{\rm tot} \approx 2.42$ signal photons per shot \cite{Gies:2017ezf}.
Even more important than the total number of attainable signal photons are their kinematics in comparison to the incident laser photons, as these decide about the principle feasibility of the measurement of the effect.
Figure \ref{fig:sketch} depicts the differential number of signal photons $\frac{{\rm d}^2N_{\rm tot}}{{\rm d}k_x{\rm d}k_z}$ as a function of the photon momenta $k_{\rm x}$ and $k_{\rm z}$. 
One can clearly see, that the signal photons (i) propagate in a direction different from the laser photons and (ii) feature a distinct energy of $\omega_{\ast} \approx 4.65$eV, 
clearly outside the spectrum of the driving lasers, which are peaked at the laser frequencies $\omega_1=\omega$ and $\omega_2=\omega_3=2\omega$, respectively.

Both effects can be easily explained: 
Energy conservation in the interaction process constrains the signal photons' main frequency to be
$\omega_{\ast} \approx \lvert \pm n_1 \omega_1 \pm n_2 \omega_2 \pm n_3 \omega_3 \rvert$, where $n_1+n_2+n_3 \overset{!}{=} 3$, for an effective interaction process of laser photons from all driving beams. 
It turns out, that momentum conservation is only compatible with one possible combination,
such that the main signal photon emission channel is given by the absorption of a photon from each of the frequency-doubled beams and the emission of a photon from the fundamental-frequency beam. As a result,
the energy of the signal photons peaks at $\omega_{\ast} \approx - \omega + 2 \omega + 2\omega=3\omega$.

\section{Summary}
\label{sec:summary}

In this article, we have discussed photonic signatures of QED vacuum nonlinearity in the collision of multiple high-intensity laser pulses in vacuum.
To this end, we have described the driving high-intensity laser fields as paraxial Gaussian beams supplemented with a Gaussian-shape pulse profile, enforcing a finite pulse duration. 
In particular with regard to the total numbers of attainable signal photons, the leading order paraxial approximation should allow for reasonably accurate predictions for high-intensity laser experiments; cf. also 
Ref.~\cite{Blinne:2018nbd} for further details. 
Our results substantiate the great potential of modern high-intensity laser systems to experimentally verify QED vacuum nonlinearities in macroscopic electromagnetic fields under well-controlled laboratory conditions for the first time.

\section*{Acknowledgments}

We are grateful to Nico Seegert and Andr\'{e} Sternbeck for many helpful discussions and
support during the development phase of the numerical algorithm.  The
work of C.K.~is funded by the Helmholtz Association through the
Helmholtz Postdoc Programme (PD-316). We acknowledge support by the
BMBF under grant No. 05P15SJFAA (FAIR-APPA-SPARC).

Computations were performed on the ``Supermicro Server 1028TR-TF'' in Jena, which
was funded by the Helmholtz Postdoc Programme (PD-316).

\end{document}